\documentclass{llncs}
\usepackage{amsfonts,amssymb,amsmath}

\advance\hoffset by -0.4in
\advance\textwidth by 0.8in
\advance\voffset by -0.3in
\advance\textheight by 0.6in
\newcommand{\CC}{\mathcal{C}}
\newcommand{\RR}{\mathcal{R}}

\newcommand{\NN}{\mathbb{N}}

\begin{document}
\title{An approach to membrane computing under
inexactitude}
\author{J. Casasnovas\and J. Mir\'o\and M. Moy\`a\and F. Rossell\'o} %
\institute{Department of Mathematics and Computer Science\\ Research 
Institute of Health Science (IUNICS)\\ University of the Balearic 
Islands\\ E-07122 Palma de Mallorca, Spain\\
\email{\{jaume.casasnovas,joe.miro,manuel.moya,cesc.rossello\}@uib.es}} %

\maketitle
\begin{abstract}	
In this paper we introduce a fuzzy version of symport/antiport
membrane systems.  Our fuzzy membrane systems handle possibly inexact
copies of reactives and their rules are endowed with threshold
functions that determine whether a rule can be applied or not to a
given set of objects, depending of the degree of accuracy of these
objects to the reactives specified in the rule.  We prove that these
fuzzy membrane systems generate exactly the recursively enumerable
finite-valued fuzzy subsets of $\mathbb{N}$.  \smallskip


\end{abstract}

\section{Introduction}

Membrane computing is a formal computational paradigm, invented in
1998 by Gh.\ P\u{a}un \cite{Paun1}, that rewrites multisets of objects
within a spatial structure inspired by the membrane structure of
living cells and according to evolution rules that are reminiscent of
the processes that take place inside cells.  Most approaches to
membrane computing developed so far have been exact: the objects used
in the computations are exact copies of the reactives involved in the
biochemical reactions modelled by the rules, and every application of
a given rule always yields exact copies of the objects it is assumed
to produce.  But, in everyday's practice, one finds that cells do not
behave in this way.  Biochemical reactions may deal with inexact,
mutated copies of the reactives involved in them, and errors may
happen when a biochemical reaction takes place.  These errors can be
due, for instance, to the inexactitude of the chemical compounds used
or to unnoticed changes in the surrounding conditions.

The inexactitude underlying cell processes made Gh.\ P\u{a}un  ask
in his very first list of open problems in membrane
computing \cite{probl1}, dated October 2000, for the development of
``approximate'' mathematical approaches.  A first answer to this
question was given by A. Obtu\l owicz and Gh.\ P\u{a}un himself by
extending the classical model to a probabilistic one \cite{ObP}.
Actually, these authors discussed several ways of introducing
probabilities in membrane computing: at the level of objects (each
object lies in a membrane with a certain probability), at the level of
rules (at each moment, each rule is fired with a certain
proba\-bility), and at the level of targets (outputs of applications
of rules are moved to each possible membrane with a certain
probability).

Beyond this probabilistic approach, Gh.\ P\u{a}un has asked more
specifically for the development of a rough set version of membrane
computing in a later list of open problems \cite{Pa-probl} and Obtu\l
owicz \cite{Obr} has discussed several possible rough set based
mathematical models of uncertainty that could be used in membrane
computing.  In the Concluding Remarks section of their aforementioned
paper \cite{ObP}, A. Obtu\l owicz and Gh.\ P\u{a}un proposed the use
of fuzzy set approaches to introduce uncertainty effects in membrane
computing.  But, although the probabilistic approaches developed in
that paper can be easily generalized to the possibilistic setting,
other fuzzy approaches to the membership of objects to membranes may
have some drawbacks.  For instance, the classical rules of fuzzy logic
---for the usual operations between fuzzy sets valued in $[0,1]$ and
involutive complement operation $c$ \cite{Klir}--- entail that, for
every membrane (including the environment) $m_{0}$, the maximum of the
membership values of a given object to all membranes other than
$m_{0}$ must be the image under $c$ of its membership value to
$m_{0}$.  Then, it is straightforward to prove that this rule implies
that, for every object, there exists some $\alpha\in [0,1]$ such that
its membership value to each membrane is $\alpha\wedge c(\alpha)$ for
all membranes but one, and $\alpha\vee c(\alpha)$ for the remaining
membrane: almost a crisp situation.

Nevertheless, there is one uncertainty aspect that cannot be handled
by means of probabilistic methods and that is suitable to being
attacked using a fuzzy set approach: the inexactitude of the reactives
involved in computations.  I.e., the fact that the actual objects used
in computations, as well as the actual output of the latter, need not
be exact copies of the reactives that are assumed to be used in the
computations or to be produced by them but only approximate copies of
these reactives.  In this paper we present a first approach to the use
of fuzzy methods to handle this kind of uncertainty.  For simplicity,
we consider only symport/antiport systems \cite{Paun2} ---membrane
systems whose rules only move reactives through membranes--- but it is
straightforward to extend our approach to other models of membrane
computing.  Our fuzzy symport/antiport systems also move objects
through membranes, but now these objects can be inexact copies of
reactives and each rule is endowed with threshold functions that
determine whether it can be applied or not to a given set of objects,
depending of the degree of approximation of these objects to the
reactives specified in the rule.  Then we prove that these fuzzy
membrane systems are universal in the sense that they generate exactly
the recursively enumerable finite-valued fuzzy subsets of $\NN$.

\section{Preliminaries}

In this section we recall some concepts on fuzzy sets and multisets
and on symport/antiport membrane systems, and we take the opportunity
to establish some notations and conventions, some of them not the
standard ones, that we shall use.

\subsection{Fuzzy sets}

Any subset $Y$ of a set $X$ can be identified with its
\emph{membership}, or \emph{characteristic}, \emph{mapping}
$\chi_{Y}:X\to \{0,1\}$, defined by $\chi_{Y}(x)=1$ if $x\in Y$ and
$\chi_{Y}(x)=0$ if $x\notin Y$.  Fuzzy subsets generalize this
interpretation of subsets as membership mappings by allowing
membership values other than 0 and 1.  Thus, a \emph{fuzzy subset} of
a set $X$ is a mapping from $X$ to the unit interval $[0,1]$.  More in
general, given any subset $I$ of $[0,1]$, an \emph{$I$-fuzzy subset}
of a set $X$ is a mapping from $X$ to $I$.  Whenever we speak about
$I$-fuzzy subsets (or multisets, see below), we shall assume that
$0,1\in I$.  A fuzzy subset of a set $X$ is \emph{finite-valued} when
its image is a finite subset of $[0,1]$, i.e., when it is $I$-fuzzy
for some finite subset $I$ of $[0,1]$.

For every fuzzy subset $\varphi:X\to [0,1]$, its \emph{$t$-level}, for every 
$t\in [0,1]$, 
is  
$$
\varphi_{t}=\{x\in X\mid \varphi(x)\geq t\}.
$$
Notice that $\varphi_{0}=X$ and, for every $t,t'\in [0,1]$, if $t\leq
t'$, then $\varphi_{t}\supseteq \varphi_{t'}$.  For every fuzzy
subsets $\varphi,\varphi':X\to [0,1]$, $\varphi=\varphi'$ if and only
if $\varphi_{t}=\varphi'_{t}$ for every $t\in [0,1]$.

If $\varphi$ is $I$-fuzzy and $t_{0},t_{1}\in I$ are such that the
open interval $]t_{0},t_{1}[$ does not contain any element of $I$,
then $\varphi_{t}=\varphi_{t_{1}}$ for every $t\in ]t_{0},t_{1}]$.
Thus, if $\varphi,\varphi'$ are $I$-fuzzy subsets of $X$, then
$\varphi=\varphi'$ if and only if $\varphi_{t}=\varphi'_{t}$ for every
$t\in I$.  This allows us, when dealing with $I$-fuzzy subsets, to
consider only their $t$-levels for $t\in I$.

A (crisp) \emph{multiset}, or \emph{bag}, over a set $V$ is simply a
mapping $d:V\to \NN$.  The usual interpretation of a multiset $d:V\to
\NN$ is that it describes a set consisting of $d(v)$ ``exact'' copies
of each $v\in V$, without specifying which element of the set is a
copy of which element of $V$.
A natural generalization of this interpretation of multisets leads to
a first definition of \emph{fuzzy multiset}, or \emph{fuzzy bag}, over
a set $V$ as a mapping $F:V\times [0,1]\to \NN$.  Such a fuzzy multiset can be
understood as describing a set consisting, for each $v\in V$ and $t\in
[0,1]$, of $F(v,t)$ ``possibly inexact'' copies of $v$ with degree of
similarity $t$ to it.  In other words, we understand that a fuzzy
multiset $F$ over $V$ describes a set endowed with a family
$(F^{(v)})_{v\in V}$ of fuzzy subsets that contains $F(v,t)$ elements $x$ such that
$F^{(v)}(x)=t$, for every $v\in V$
and $t\in [0,1]$.

It will be convenient for our purposes to take a slightly modified
definition of fuzzy multiset.  The first modification affects both
crisp and fuzzy multisets.  Sometimes we shall need to represent the
fact that the set described by a multiset (resp., a fuzzy multiset)
$F$ over $V$ contains an arbitrarily large number of copies of some
elements $v$ of $V$ (resp., with some degree of similarity $t$).  We
shall do it by writing $F(v)=\infty$ (resp., $F(v,t)=\infty$).  Thus,
our (crisp and fuzzy) multisets will actually take values in
$\NN\cup\{\infty\}$.  To simplify the notations, from now on we shall
denote this set $\NN\cup\{\infty\}$ by $\NN_{\infty}$.

On the other hand, we impose two limitations on the interpretation of
a fuzzy multiset as a set that allow us to modify its
definition; cf.\ \cite{CR-fb}. First, we shall assume that if an element of
the set described by a fuzzy multiset over $V$ is an inexact copy of
$v\in V$ with degree of similarity $t>0$, then it cannot be an inexact
copy of any other element in $V$ with any non-zero degree of
similarity.  And second, we shall also assume that the set described
by a fuzzy multiset over $V$ does not contain any element that is not
a copy of some $v\in V$ with some non-zero degree of similarity, or,
rather, we shall not take into consideration these elements.  These
conditions entail that, for every $v\in V$, the value $F(v,0)$ must be
equal to $\sum_{w\in V-\{v\}}\sum_{t\in ]0,1]} F(w,t)$ and in
particular that the restriction of $F$ to $V\times \{0\}$ is
determined by the restriction of $F$ to $V\times ]0,1]$.

These restraints allow us to define in this paper a \emph{fuzzy
multiset} over a set $V$ as a mapping
$$
F:V\times ]0,1]\to \NN_{\infty}.
$$
Not having to care about the images under fuzzy multisets of the
elements of the form $(v,0)$ will greatly simplify some of the
definitions and results that will be introduced in the main body of
this paper.

For every $I\subseteq [0,1]$, we shall denote $I-\{0\}$ by $I^{+}$.
Consistently with our definition of fuzzy multiset over a set $V$,
given any $I\subseteq [0,1]$, an \emph{$I$-fuzzy multiset over $V$}
will be a mapping $F:V\times I^+\to \NN_{\infty}$.  Every such
$I$-fuzzy multiset can be understood as defined on the whole $V\times
]0,1]$ by extending it by means of $F(v,t)=0$ for every $v\in V$ and
$t\in ]0,1]-I^+$.

A fuzzy multiset $F:V\times ]0,1]\to \NN_{\infty}$ over a
\emph{finite} set $V$ is \emph{finite-valued} when $F(v,t)=0$ for all
ordered pairs $(v,t)\in V\times ]0,1]$ except a finite number of them.
This is equivalent to say that the image of $F$ is finite and the 
preimage under $F$ of every $n\in \NN_{\infty}-\{0\}$ is a finite set. 
If $I\subseteq [0,1]$ is a finite set such that $F(v,t)\neq 0$ for
some $v\in V$ implies $t\in I$, then we shall identify such a fuzzy
multiset with the $I$-fuzzy multiset $F:V\times I^+\to
\NN_{\infty}$ obtained as its restriction to $V\times I^+$.

\subsection{Crisp symport/antiport membrane systems}

In this subsection we explain in detail the basic model of membrane
systems with symport/anti\-port rules, with notations that are not the
usual ones but that will be helpful in the generalization of this
model to the fuzzy setting.  The interested reader can look up Chapter
4 of Gh.  P\u{a}un's textbook on molecular computing \cite{Paun3} and
the references cited therein for more information on this model of
computation.

Given an alphabet $V$, we denote by $V^{*}$ the set of words over $V$.
Given a word $w\in V^{*}$, we denote by $|w|$ the length of $w$ and,
given a letter $a\in V$, by $|w|_{a}$ the number of occurrences of $a$
in $w$.

A \emph{membrane structure} $\mu$ is a finite rooted tree whose nodes
are called \emph{membranes}. We shall always denote by $M$ the set of
membranes of a membrane structure and, in practice, we shall assume
that these membranes are injectively labelled by natural numbers, in
such a way that the root's label is $1$.  The edges of
$\mu$ are oriented pointing to the root.

This tree represents a hierarchical structure of nested membranes,
with the edges representing the relation ``being directly inside'': an
edge going from a membrane $m$ to a membrane $m'$ means that $m$ is
directly included in $m'$.  The tree's root $1$ is then called the
\emph{skin membrane}, because it surrounds the whole membrane system,
and the tree's leaves are called \emph{elementary membranes}, because
no further membrane lies inside them.

We expand every membrane structure $\mu$ by adding a new node to it
labelled $env$ and an arc going from 1 to $env$; let $\overline{\mu}$
denote the resulting tree and $\overline{M}$ its set of nodes
$M\cup\{env\}$.  This new node $env$ is called the \emph{environment},
because it surrounds the skin membrane.  In this way, $env$ becomes
the root of $\overline{\mu}$.  Although, formally, $env$ is not a
membrane, when we generically talk about membranes, we shall
include it unless we explicitly state otherwise.

For every $m\in M$, we shall denote by $\varepsilon(m)$ the target
node of the arc in $\overline{\mu}$ whose source node is $m$, i.e., the
membrane which $m$ is directly included into.  Notice that
$\varepsilon(1)=env$, and that if $m\neq 1$, then $\varepsilon(m)\in
M$.

We understand that every $m\in \overline{M}$ defines a \emph{region}
$K_{m}$.  For an elementary membrane, it would represent the space
enclosed by it, and for any other membrane it would represent the
space comprised between this membrane and those directly included in
it.  The $env$ node also defines a region $K_{env}$, which would represent
the space outside the skin membrane: the \emph{environment}, indeed.
The rules in the membrane systems considered in this paper will move
in a controlled way objects from regions $K_{m}$ to regions
$K_{\varepsilon(m)}$ and, the other way round, from regions
$K_{\varepsilon(m)}$ to regions $K_{m}$.

At each moment, every such region contains a set of objects that are
copies of elements of a certain finite set of \emph{reactives}.  If
$V$ stands for this set of reactives, the content of all regions
$K_{m}$, with $m\in \overline{M}$, at any moment is represented by means
of an $\overline{M}$-indexed family $(F_{m})_{m\in \overline{M}}$ of
{multisets} over $V$
$$
F_{m}:V\to \NN_{\infty},\quad m\in \overline{M}.
$$
We shall call such an $\overline{M}$-indexed family
of {multisets} $(F_{m})_{m\in \overline{M}}$ a \emph{configuration}.

Now, a \emph{symport/antiport membrane system},  a
\emph{P-system} for short, is a structure
$$
\Pi=(V, V_{out}, \mu, m_{out}, (S_{m})_{m\in \overline{M}}, (\mathcal{R}_{m})_{m\in M})
$$
where:
\begin{itemize}

\item $V$ is the set of \emph{reactives} used by the membrane system; 
it is a finite set.

\item $V_{out}\subseteq V$ is the set of \emph{output reactives};
these are the only objects that matter at the end of a
computation.\footnote{In the original definition of symport/antiport
membrane systems, no set of output reactives is distinguished, i.e.,
$V=V_{out}$.  But, the specification of a set of output reactives will
simplify a proof in the fuzzy setting, and it does not
increase the computational power of these systems: see Remark 
\ref{theremark} at the end of this section.}

\item $\mu$ is a membrane structure, with set of membranes $M$.

\item $m_{out}\in M$ is the \emph{output membrane}; the results of 
the computations are read in the region defined by this membrane.

\item $(S_{m})_{m\in \overline{M}}$ is a configuration, called
\emph{initial}, that describes the initial content of each region
$K_{m}$.  We assume that, for each $v\in V$,
$S_{env}(v)$ is either 0 or $\infty$, and that $S_{m}(v)\neq\infty$
for every $m\in M$ and $v\in V$.

\item For every $m\in M$, $\mathcal{R}_{m}$ is a finite set of
\emph{evolution rules} associated to the membrane $m$: notice that the
environment has no rule associated to it.  These rules represent
changes of place of objects between $K_{m}$ and regions adjacent to
it.  Every rule in $\mathcal{R}_{m}$ has the form
$$
R=(\underline{a},in; \underline{b},out),
$$
where $\underline{a}, \underline{b}\in V^{*}$ represent the multisets
of reactives that enter ($in$) or exit ($out$) the region defined by
the membrane $m$ under the action of this rule.  

When $\underline{b}$ or $\underline{a}$ is the empty word $\lambda$,
the rule $R$ is said to be a \emph{symport rule}, and it is simply
written  $(\underline{a},in)$ or $(\underline{b},out)$,
respectively.  When $\underline{a},\underline{b}\neq \lambda$, the
rule $R$ is said to be an \emph{antiport rule}. 
\end{itemize}

Let $m_{0}\in M$ be a membrane and $\varepsilon(m_{0})$ the membrane
in $\overline{\mu}$ directly outside it.  An evolution rule
$R=(\underline{a},in; \underline{b},out) \in \mathcal{R}_{m_{0}}$
\emph{can be applied} to a configuration 
$(F_{m})_{m\in \overline{M}}$ when,
for every $v\in V$,
$$
F_{\varepsilon(m_{0})}(v)\geq |\underline{a}|_{v}\mbox{ and }
F_{m_{0}}(v)\geq |\underline{b}|_{v}.
$$
And when it can be applied, its \emph{application} produces a new
configuration $(F'_{m})_{m\in \overline{M}}$, which we call the
\emph{result} of this application, that is obtained as follows:
\begin{itemize}
\item $F'_{m}=F_{m}$ if $m\neq m_{0},\varepsilon(m_{0})$;

\item $F'_{m_{0}}(v)=F_{m_{0}}(v)-|\underline{b}|_{v}+|\underline{a}|_{v}$ for every
$v\in V$;

\item $F'_{\varepsilon(m_{0})}(v)=F_{\varepsilon(m_{0})}(v)+
|\underline{b}|_{v}-|\underline{a}|_{v}$ for every $v\in V$.
\end{itemize}

This represents that $(F'_{m})_{m\in \overline{M}}$ is obtained from
$(F_{m})_{m\in \overline{M}}$ by moving, for every $v\in V$,
$|\underline{a}|_{v}$ copies of $v$ from $K_{\varepsilon(m_{0})}$ to
$K_{m_{0}}$ and $|\underline{b}|_{v}$ copies of $v$ from $K_{m_{0}}$
to $K_{\varepsilon(m_{0})}$.  

A \emph{transition} for a P-system $\Pi$ consists of a maximal
simultaneous application of evolution rules.  These rules are chosen
non-deterministically in such a way that  no further rule
in $\mathcal{R}_{m}$, for any $m$, can be triggered simultaneously to them.
Formally, a transition consists of the simultaneous application to a
configuration $(F_{m})_{m\in \overline{M}}$ of a family of rules
$$
(R_{m,1},\ldots,R_{m,r_{m}})_{m\in M},
$$
with
$R_{m,i}=(\underline{a_{m,i}},in;\underline{b_{m,i}},out)\in \RR_{m}$, for every $m\in 
M$ and $i=1,\ldots,r_{m}$. This family of rules 
must satisfy the following two conditions:
\begin{itemize}
\item[(1)]  For every elementary membrane $m$ and for every $v\in V$,
$$
\sum_{i=1}^{r_{m}} |\underline{b_{m,i}}|_{v}\leq F_{m}(v);
$$
for every non-elementary membrane $m\in M$, say with 
$m=\varepsilon(m_{1})=\ldots=\varepsilon(m_{l})$, and for every $v\in V$,
$$
\sum_{j=1}^{l}\sum_{i=1}^{r_{m_{j}}} |\underline{a_{m_{j},i}}|_{v}+
\sum_{i=1}^{r_{m}} |\underline{b_{m,i}}|_{v}\leq F_{m}(v);
$$
and, finally,
$$
\sum_{i=1}^{r_{1}} |\underline{a_{1,i}}|_{v}\leq F_{env}(v).
$$

\item[(2)] No further rule can be added to any 
$(R_{m,1},\ldots,R_{m,r_{m}})$ in such a way that the resulting family 
of rules still satisfies the application condition  (1).
\end{itemize}
These conditions globally impose that there are enough reactives in
all regions to allow the simultaneous application of all rules
$R_{m,i}$, but that there are not enough reactives to allow the
application of any further rule.

Now, the simultaneous application of these rules to a
configuration $(F_{m})_{m\in \overline{M}}$ produces a new
configuration $(\widehat{F}_{m})_{m\in \overline{M}}$ that is obtained
as follows:
\begin{itemize}
\item For every elementary membrane $m$ and for every $v\in V$,
$$
\widehat{F}_{m}(v)=F_{m}(v)+\sum_{i=1}^{r_{m}}
|\underline{a_{m,i}}|_{v}-\sum_{i=1}^{r_{m}}
|\underline{b_{m,i}}|_{v}.
$$

\item For every non-elementary membrane $m\in M$, say with 
$m=\varepsilon(m_{1})=\ldots=\varepsilon(m_{l})$, and for every $v\in V$,
$$
\widehat{F}_{m}(v)=F_{m}(v)+ \sum_{i=1}^{r_{m}}
|\underline{a_{m,i}}|_{v}+ \sum_{j=1}^{l}\sum_{i=1}^{r_{m_{j}}}
|\underline{b_{m_{j},i}}|_{v}-\sum_{j=1}^{l}\sum_{i=1}^{r_{m_{j}}}
|\underline{a_{m_{j},i}}|_{v}- \sum_{i=1}^{r_{m}}
|\underline{b_{m,i}}|_{v}.
$$

\item And, for every $v\in V$,
$$
\widehat{F}_{env}(v)=F_{env}(v)+ \sum_{i=1}^{r_{1}}
|\underline{b_{1,i}}|_{v}-\sum_{i=1}^{r_{1}}
|\underline{a_{1,i}}|_{v}.
$$
\end{itemize}

We shall forbid the existence in $\RR_{1}$ of any symport rule of the form
$(\underline{a},in)$ with $\underline{a}\in V^{*}$ such that
$S_{env}(v)=\infty$ if $|\underline{a}|_{v}>0$, because any such rule
could be applied an infinite number of times in any transition.

A finite sequence of transitions between configurations of a 
P-system $\Pi$, starting with the initial configuration, is
called a \emph{computation} with respect to $\Pi$.  A computation $C$
\emph{halts} when it reaches a \emph{halting configuration}
$(H(C)_{m})_{m\in \overline{M}}$ where no rule can be applied.  The
\emph{output} of such a halting computation $C$ is the final number of
output reactives contained in the  region defined by the output 
membrane:
$$
Out_{\Pi,C}=\sum_{v\in V_{out}} H(C)_{m_{out}}(v).
$$
A computation that does not halt does not yield any output.

The set $Gen(\Pi)\subseteq\NN$ \emph{generated} by $\Pi$ is the set of
all outputs $Out_{\Pi,C}$ of halting computations $C$ with respect to $\Pi$.

We have now the following result; see \cite{Paun2,Paun3,MPP}.

\begin{theorem}\label{crisp}
A subset of $\NN$ is recursively enumerable if and only if it is
generated by some P-system.

Moreover, every recursively enumerable subset of $\NN$ can be
generated by a P-system that satisfies the following conditions:
\begin{itemize}
\item its membrane structure has only two nodes, and the output membrane is the
elementary one;

\item it has only symport rules; 

\item all reactives used by the P-system are output reactives;

\item the rules associated to the output membrane are $(\alpha,in)$, 
$(\#,in)$ and $(\#,out)$ for some specific reactives $\alpha$ and 
$\#$, and $\alpha$ is the only reactive that may enter the output 
membrane in any halting computation;

\item in the initial configuration, both the skin membrane and the
output membrane do not contain any copy of this reactive $\alpha$.  $\Box$
\end{itemize}
\end{theorem}

\begin{remark}\label{theremark}
	Notice that last theorem establishes that every recursively enumerable 
	subset of $\NN$ is generated by a P-system all whose reactives are 
	considered as output reactives. Since, by Church-Turing Thesis, a 
	P-system with a set of output reactives specified will generate a recursively enumerable 
	subset of $\NN$, we deduce that the specification of a set of output 
	reactives does not increase the computational power of the model, as 
	we claimed when we defined our P-systems.
	
	Besides, we also have that every recursively enumerable subset of $\NN$ is
	generated by a P-system with only one output reactive: the only
	reactive $\alpha$ that may enter the output membrane in any halting
	computation in the P-system given by the last theorem.
\end{remark}

\section{The fuzzy model}

We assume henceforth the existence of a \emph{universe} $X$
containing all objects we use in computations.

Roughly described, a fuzzy P-system will be a structure similar to a
crisp P-system, supported on a membrane structure that defines regions
whose contents evolve following rules that specify the transport of
reactives through membranes.  But the details will be quite different.

To begin with, we shall use \emph{reactives} as ``ideal definitions''
of chemical compounds, and hence they are fuzzy subsets of $X$: for
every reactive $v:X\to [0,1]$, we understand that $v(x)=t$ denotes that
the object $x\in X$ is a copy of  $v\in V$ with a degree
$t$ of exactitude.  So, $v(x)=1$ means that $x$ is an \emph{exact
copy} of the reactive $v$, and $v(x)=0$ means that $x$ cannot
represent in any way the reactive $v$.

Actually, every reactive will be, for the purposes of each fuzzy
P-system, a finite-valued fuzzy subset of $X$: this represents that,
in any fuzzy P-system, only a finite set of values of accuracy of
objects to reactives will be taken into account.  This can be seen as
translating nature's discreteness, or that the accuracy of an object
to a reactive cannot be measured exactly, but only up to some
threshold.  As we shall explain in the
Conclusion, this finite-valuedness assumption does not decrease the
computational power of our fuzzy P-systems: if we allowed the
reactives to take values in the whole $[0,1]$, the set of natural
numbers generated by a fuzzy P-system would still be finite-valued.

We shall say that an object $x\in X$ is \emph{similar} to a reactive
$v\in V$ when $v(x)>0$.  To simplify the definition of an application
of a rule, and as it was already hinted in \S 2.1, we shall assume in
this paper that each object in $X$ is similar to at most one reactive,
and it will be clear from the  definition of the
application of a rule that in each fuzzy P-system we shall not care
about objects that are not similar to some reactive among those used
in it.

As in the crisp case, fuzzy P-systems will be supported by a membrane
structure and each membrane in it will define a region.  But, the
reactives being fuzzy sets, the content of these regions at each
moment will be formally described by means of an
$\overline{M}$-indexed family of \emph{fuzzy multisets} over a set $V$
of reactives.  These fuzzy multisets specify, for every $v\in V$ and
for every value $t\in ]0,1]$, how many objects  in each
region $K_{m}$  are copies of the reactive $v$ with degree of
accuracy~$t$.

Since each fuzzy P-system will involve only a finite set of reactives
$V$, and, for the purposes of each specific P-system, we consider each
reactive as a finite-valued fuzzy subset of $X$, all possible values
of accuracy of objects to reactives used in a given fuzzy P-system
form a finite subset of $[0,1]$.  Thus, we shall specify in the
description of a fuzzy P-system a finite subset $I$ of $[0,1]$ that
will contain all these images as well as all other elements in $[0,1]$
needed in that description.  Then, a \emph{configuration} for this
fuzzy P-system, with set of membranes $M$ and set of reactives $V$,
will be a family of $I$-valued fuzzy multisets $(F_{m})_{m\in
\overline{M}}$ over $V$,
$$
F_{m}: V\times I^+\to \NN_{\infty},\qquad m\in \overline{M}.
$$
Each such mapping $F_{m}$ specifies, for every $v\in V$ and for every
$t\in I^{+}$, how many objects there exist in the region $K_{m}$ such
that $v(x)= t$ at the moment described by the configuration.  We
impose several conditions on these configurations.  First, every
$F_{m}$ with $m\in M$ is such that $F_{m}(v,t)<\infty$ for every $v\in
V$ and for every $t\in I^{+}$: this translates the fact that the
regions defined by the membranes other than the environment can only
contain at any time a finite set of objects.  Second, 
we allow the environment to contain at every
moment an unlimited supply of copies of some reactives $v$, and then
with all possible degrees of accuracy $t\in I^+$: we shall represent
it by writing $F_{env}(v,t)=\infty$ for every $t\in I^+$.

Now, a \emph{fuzzy symport/antiport membrane system},  a 
\emph{fuzzy P-system} for short, is a structure
$$
\Pi=(V, V_{0},\mu, m_{out}, I, (S_{m})_{m\in \overline{M}}, 
(\mathcal{R}_{m})_{m\in M}),
$$
where:
\begin{itemize}

\item $V$ is the finite set of \emph{reactives} used by the membrane system.

\item $V_{out}\subseteq V$ is the set of \emph{output reactives}.

\item $\mu$ is a membrane structure, with set of membranes $M$.

\item $m_{out}\in M$ is the \emph{output membrane}. 

\item $I$ is a finite subset of $[0,1]$ containing 0 and 1.

\item $(S_{m})_{m\in \overline{M}}$ is a family of $I$-valued fuzzy
multisets over $V$, called the \emph{initial configuration}, which
describes the initial content of all regions $K_{m}$.  We impose that,
for each $v\in V$,  either   $S_{env}(v,t)=0$ for every
$t\in I^{+}$ or  $S_{env}(v,t)=\infty$ for every $t\in I^{+}$.
This translates the assumption that, for every reactive $v$, it either
happens that the environment does not contain any object similar to it
or that it contains an unbounded homogeneous supply of copies of it.

\item For every  $m\in M$, $\mathcal{R}_{m}$ is a finite set
of \emph{evolution rules} associated to $m$.  
Each evolution rule in $\mathcal{R}_{m}$ has the form
$$
R=((\underline{a},in; \underline{b},out),\tau_{in},\tau_{out}),
$$
where:
\begin{itemize}
\item[---] $(\underline{a},in; \underline{b},out)$ is a crisp
symport/antiport rule; we shall say that a
reactive $v$ is \emph{incoming} (resp., \emph{outgoing}) for this
rule $R$ when $|\underline{a}|_{v}>0$ (resp.,
$|\underline{b}|_{v}>0$).

\item[---] $\tau_{in},\tau_{out}: V\to I$ are \emph{threshold}
functions that determine, for every 
incoming or outgoing reactive for $R$, respectively,
the degree of accuracy of an object to this reactive that is necessary for
this object to be considered as this reactive to the effect of
triggering an application of this rule.

We impose on these threshold functions that $\tau_{in}(v)>0$ for every
incoming reactive and $\tau_{out}(v)>0$ for every outgoing reactive: 
objects that are not similar to an incoming or outgoing reactive 
can never play its role in the application of a rule.
Moreover, and for simplicity, we do not impose any threshold condition on
reactives that are not incoming or outgoing: if $|\underline{a}|_{v}=0$,
then $\tau_{in}(v)=0$, and if $|\underline{b}|_{v}=0$, then
$\tau_{out}(v)=0$.
\end{itemize}

As in the crisp case, when $\underline{b}$ or $\underline{a}$ is the
empty word $\lambda$, $R$ is said to be a \emph{symport rule}, and we
shall simply write it as $((\underline{a},in),\tau)$ or
$((\underline{b},out),\tau)$, respectively: in the rules of the first
type, $\tau$ represents $\tau_{in}$, and in those of the second type,
it represents $\tau_{out}$.  When $\underline{a},\underline{b}\neq
\lambda$, $R$ is said to be an \emph{antiport rule}.

Also as in the crisp case, and for the very same reason as then, we
forbid the existence in $\RR_{1}$ of symport rules of the form
$((\underline{a},in),\tau)$ with $\underline{a}\in V^{*}$ such that
$S_{env}(v,-)=\infty$ if $|\underline{a}|_{v}>0$.

\end{itemize}

Let $m_{0}\in M$ be a membrane and $\varepsilon(m_{0})$ the membrane 
in $\overline{\mu}$ directly outside it.
An evolution rule
$$
R=((\underline{a},in; \underline{b},out),\tau_{in},\tau_{out})
$$
in $\mathcal{R}_{m_{0}}$ \emph{can be triggered} in a configuration 
$(F_{m})_{m\in \overline{M}}$ when,
for every $v\in V$,
$$
\sum_{t\geq \tau_{in}(v)}F_{\varepsilon(m_{0})}(v,t)\geq 
|\underline{a}|_{v}\mbox{ and }
\sum_{t\geq \tau_{out}(v)}F_{m_{0}}(v,t)\geq |\underline{b}|_{v}.
$$
This means that  there are more copies of every
incoming or outgoing reactive in the regions $K_{\varepsilon(m_{0})}$ 
and $K_{m_{0}}$, respectively, within the degree of accuracy
required by the threshold functions, than the specified quantities.

When a rule 
$$
R=((\underline{a},in; \underline{b},out),\tau_{in},\tau_{out})\in \mathcal{R}_{m_{0}}
$$
can be triggered in a configuration $(F_{m})_{m\in \overline{M}}$, an
\emph{application} of it modifies this configuration into a new
configuration $(F'_{m})_{m\in \overline{M}}$, which we call the
\emph{result} of this specific application.  This new configuration is
obtained as follows:

\begin{enumerate}
\item[(1)] 
For every reactive $v\in V$, we choose $|\underline{a}|_{v}$ objects in
$K_{\varepsilon(m_{0})}$ with degree of accuracy to $v$ at least
$\tau_{in}(v)$.  Formally, to do it, for every $v\in V$, we take a 
mapping $enter^{R}_{v}:I^{+}\to \NN$ such that:
\begin{itemize}
\item[---] If $t<\tau_{in}(v)$, then $enter^{R}_{v}(t)=0$. 

\item[---] If $t\geq \tau_{in}(v)$, then $0\leq enter^{R}_{v}(t)\leq
F_{\varepsilon(m_{0})}(v,t)$.

\item[---] $\sum_{t\in I^{+}} enter^{R}_{v}(t)=|\underline{a}|_{v}$.
\end{itemize}
Notice in particular that if $|\underline{a}|_{v}=0$, then $enter^{R}_{v}(t)=0$ 
for every $t\in I^{+}$.

This corresponds to choosing, for every $t\geq \tau_{in}(v)$, a
certain number $enter^{R}_{v}(t)$ of objects $x$ in
$K_{\varepsilon(m_{0})}$ such that $v(x)=t$ and in such a way that the
total amount of these objects is $|\underline{a}|_{v}$.  These
objects, or, rather, the number of them within each degree $t\geq
\tau_{in}(v)$ of accuracy to $v$, are chosen in a non-deterministic
way: taking a different mapping $enter^{R}_{v}$ would correspond to a
different application of the rule and hence it could lead to a
different result.

\item[(2)] In a similar way, for every $v\in V$, we choose
$|\underline{b}|_{v}$ objects in $K_{m_{0}}$ with degree of accuracy
to $v$ at least $\tau_{out}(v)$.  As before, we do it by taking, for
every $v\in V$, a mapping $exit^{R}_{v}:I^{+}\to \NN$ such that:
\begin{itemize}
\item[---] If $t<\tau_{out}(v)$, then $exit^{R}_{v}(t)=0$. 

\item[---] If $t\geq \tau_{out}(v)$, then $0\leq exit^{R}_{v}(t)\leq
F_{m_{0}}(v,t)$.

\item[---] $\sum_{t\in I^{+}} exit^{R}_{v}(t)=|\underline{b}|_{v}$.
\end{itemize}
We have again that if $|\underline{b}|_{v}=0$, then $exit^{R}_{v}(t)=0$ for every $t\in
I^{+}$.

\item[(3)]
For every reactive $v\in V$, we move from $K_{m_{0}}$ to
$K_{\varepsilon(m_{0})}$ the $|\underline{b}|_{v}$ possibly inexact
copies of it that have been chosen by means of the mapping
$exit^{R}_{v}$, and we move from $K_{\varepsilon(m_{0})}$ to
$K_{m_{0}}$ the $|\underline{a}|_{v}$ possibly inexact  copies of it that have been
chosen by means of $enter^{R}_{v}$.  This leads to a new
configuration $(F'_{m})_{m\in \overline{M}}$ defined as follows:

\begin{itemize}
\item $F'_{m}=F_{m}$ if $m\neq m_{0},\varepsilon(m_{0})$.

\item $F'_{m_{0}}(v,t)=F_{m_{0}}(v,t)-exit^{R}_{v}(t)+enter^{R}_{v}(t)$ for every
$v\in V$ and $t\in I^{+}$.

\item  $F'_{\varepsilon(m_{0})}(v,t)=F_{\varepsilon(m_{0})}(v,t)+exit^{R}_{v}(t)-enter^{R}_{v}(t)$ for every 
$v\in V$ and $t\in I^{+}$.
\end{itemize}
Consequently, if $F_{env}(v,t)=\infty$, then
$F'_{env}(v,t)=\infty$, too, and if $F_{env}(v,t)\neq\infty$, then
$F'_{env}(v,t)\neq\infty$ either.
\end{enumerate}

This new configuration $(F'_{m})_{m\in \overline{M}}$ is the
\emph{result} of this application of $R$.  Let us point out again that
a given rule may admit several applications to a given configuration,
yielding different results, depending on the mappings taken in steps
(1) and (2).  This does not happen in the crisp case.

Now, a \emph{transition} for a fuzzy P-system $\Pi$ consists of a
maximal simultaneous application of rules in the same sense as in the
crisp case: the triggering condition must be satisfied simultaneously
for all rules, and then all steps (1) and (2) corresponding to rules
being applied in one transition are performed simultaneously, and
finally all steps (3) are performed together.  The rules applied in
a given transition are chosen  non-deterministically  but so that
no further rule in $\mathcal{R}_{m}$ for any $m$ can be triggered
simultaneously to them.  In particular, a given rule can be triggered
several times in the same transition, provided enough copies of the
corresponding incoming and outgoing reactives are available within the
required degree of exactitude.

Formally, a transition consists of the simultaneous application to a
configuration $(F_{m})_{m\in \overline{M}}$ of a family of rules
$$
(R_{m,1},\ldots,R_{m,r_{m}})_{m\in M},
$$
with
$$
R_{m,i}=((\underline{a_{m,i}},in;\underline{b_{m,i}},out),\tau^{m,i}_{in},
\tau^{m,i}_{out}) \in \RR_{m},\quad m\in M,\ i=1,\ldots,r_{m}.
$$
These rules must satisfy that:
\begin{enumerate}
\item[(a)] For every $v\in V$ and for every $t\in I^+$,
\begin{itemize}
\item  for every elementary membrane $m$,
$$
\sum_{i\mathrm{\, s.t.\, }\tau^{m,i}_{out}(v)\geq t} 
|\underline{b_{m,i}}|_{v}\leq \sum_{t'\geq t} F_{m}(v,t');
$$

\item for every non-elementary membrane $m\in M$  with 
$m=\varepsilon(m_{1})=\ldots=\varepsilon(m_{l})$,
$$
\sum_{i\mathrm{\, s.t.\, }\tau^{m,i}_{out}(v)\geq t} |\underline{b_{m,i}}|_{v}
+
\sum_{j=1}^{l}\,\sum_{i\mathrm{\,  s.t.\, }\tau^{m_{j},i}_{in}(v)\geq t}  
|\underline{a_{m_{j},i}}|_{v} \!\leq\!
\sum_{t'\geq t} F_{m}(v,t');
$$

\item finally, as far as $env$ goes,
$$
\sum_{i\mathrm{\,  s.t.\, }\tau^{1,i}_{in}(v)\geq t} |\underline{a_{1,i}}|_{v}\leq 
\sum_{t'\geq t} F_{env}(v,t').
$$
\end{itemize}

\item[(b)] No further rule can be added to any
$(R_{m,1},\ldots,R_{m,r_{m}})$ so that the resulting family of rules
still satisfies condition (a).
\end{enumerate}
And then the simultaneous application of these rules to a
configuration $(F_{m})_{m\in \overline{M}}$ produces a new
configuration $(\widehat{F}_{m})_{m\in \overline{M}}$ that is obtained
as follows:

\begin{enumerate}
\item[(c)] For every  $m\in M$, for every rule $R_{m,i}$, 
$i=1,\ldots,r_{m}$, and for every $v\in V$, we take mappings
$enter^{m,i}_{v},exit^{m,i}_{v}:I^{+}\to \NN$ in such a way that:
\begin{itemize}
\item  $\sum_{t\in I^{+}} enter^{m,i}_{v}(t)=|\underline{a_{m,i}}|_{v}$
and $\sum_{t\in I^{+}} exit^{m,i}_{v}(t)=|\underline{b_{m,i}}|_{v}$.

\item  If $t<\tau^{m,i}_{in}(v)$, then $enter^{m,i}_{v}(t)=0$, 
and if  $t<\tau^{m,i}_{out}(v)$, then $exit^{m,i}_{v}(t)=0$.

\item  For every $t\in I^+$, 
\begin{itemize}
\item  for every elementary membrane $m$,
$$
\sum_{i=1}^{r_{m}} exit^{m,i}_{v}(t)\leq F_{m}(v,t);
$$

\item for every non-elementary membrane $m\in M$  with 
$m=\varepsilon(m_{1})=\ldots=\varepsilon(m_{l})$,
$$
\sum_{i=1}^{r_{m}} exit^{m,i}_{v}(t) +
\sum_{j=1}^{l}\sum_{i=1}^{r_{m_{j}}} enter^{m_{j},i}_{v}(t)\leq F_{m}(v,t);
$$

\item $\displaystyle \sum_{i=1}^{r_{1}} enter^{1,i}_{v}(t)\leq
F_{env}(v,t).$
\end{itemize}
\end{itemize}

\item[(d)] The new configuration $(\widehat{F}_{m})_{m\in 
\overline{M}}$ produced by this application is obtained as follows: 
for every $v\in V$ and $t\in I^+$,

\begin{itemize} 
		
\item for every elementary membrane $m$,
$$
\widehat{F}_{m}(v,t)=F_{m}(v,t)+\sum_{i=1}^{r_{m}}
enter^{m,i}_{v}(t)-\sum_{i=1}^{r_{m}} exit^{m,i}_{v}(t);
$$

\item for every non-elementary membrane $m\in M$  with 
$m=\varepsilon(m_{1})=\ldots=\varepsilon(m_{l})$, 
$$
\begin{array}{rl}
\widehat{F}_{m}(v,t)=F_{m}(v,t) & + \sum_{i=1}^{r_{m}}
enter^{m,i}_{v}(t) +\sum_{j=1}^{l}\sum_{i=1}^{r_{m_{j}}}
exit^{m_{j},i}_{v}(t)\\ & -\sum_{i=1}^{r_{m}} exit^{m,i}_{v}(t)
-\sum_{j=1}^{l}\sum_{i=1}^{r_{m_{j}}} enter^{m_{j},i}_{v}(t);
\end{array}
$$

\item Finally,
$$
\widehat{F}_{env}(v,t)=F_{env}(v,t)+\sum_{i=1}^{r_{1}} exit^{1,i}_{v}(t)
-\sum_{i=1}^{r_{1}} enter^{1,i}_{v}(t).
$$
\end{itemize}
\end{enumerate}

A finite sequence of transitions between configurations of a fuzzy
P-system $\Pi$, starting with the initial configuration, is
called a \emph{computation} with respect to $\Pi$.  A computation 
\emph{halts} when it reaches a \emph{halting configuration}
where no rule can be triggered.

Given a halting computation $C$ with halting configuration
$(H(C)_{m})_{m\in \overline{M}}$, the (crisp) multiset over $I^+$
\emph{associated} to it is 
$$
\begin{array}{rrcl}
H_{C}: & I^{+} & \to & \NN \\
 & t & \mapsto & \sum_{v\in V_{0}} H(C)_{m_{out}}(v,t)
 \end{array}
$$
Thus, for every $t\in I^{+}$, $H_{C}(t)$ is the number of objects in
the output region that, at the end of the computation, are copies of
some output reactive with degree of exactitude $t$.

Then, the \emph{output} of a halting computation $C$ will be the fuzzy 
subset of $\NN$ 
$$
\begin{array}{rrcl}
	Out_{\Pi,C}: & \NN & \to & I\\
	& n & \mapsto & \bigvee\{t\mid H_{C}(t)= n\}
\end{array}
$$
In words, $Out_{\Pi,C}(n)$ is the greatest degree of exactitude $t$ in
$I$ for which, at the end of the computation $C$, there exist $n$
objects in the output region that are copies of some output reactive with
degree of exactitude $t$.

Finally, the fuzzy set of natural numbers \emph{generated} by a fuzzy
membrane system $\Pi$ is the join of all the outputs of halting
computations with respect to $\Pi$.  This is the mapping
$Gen_{\Pi}:\NN\to I$ defined by
$$
Gen_{\Pi}(n)=\bigvee_{C\ \mathrm{halting}} Out_{\Pi,C}(n),\qquad n\in \NN.
$$
Thus,
$$
\begin{array}{rl}
Gen_{\Pi}(n) & =\bigvee \Bigl\{\bigvee\{t\in I^+\mid H_{C}(t)= n\}\mid C\ 
\mathrm{halting}\Bigr\}\\[2ex]
& =\bigvee \{t\in I^+\mid H_{C}(t)= n \mbox{ for some
halting computation $C$}\}.
\end{array}
$$
Notice that, $I$ being finite, this supremum is actually a
maximum, and that if $H_{C}(t)\neq n$ for every halting computation
$C$, then $Gen_{\Pi}(n)=\bigvee \emptyset=0$.

The following lemma is a direct consequence of the last description of
$Gen_{\Pi}(n)$ and the finiteness of $I$.

\begin{lemma}\label{lemaprevi1}
For every fuzzy P-system $\Pi$, for every $n\in \NN$ and for every
$t_{0}\in I^+$, $Gen_{\Pi}(n)\geq t_{0}$ if and only if there exists some
halting computation $C$ and some $t\geq t_{0}$ such that $H_{C}(t)= 
n$.   $\Box$
\end{lemma}

Crisp P-systems can be seen as special cases of their fuzzy version.
Indeed, every crisp P-system
$$
\Pi=(V, V_{out},\mu, m_{out}, (S_{m})_{m\in \overline{M}},
(\mathcal{R}_{m})_{m\in M})
$$
defines a fuzzy P-system
$$
\Pi^{f}=(V, V_{out},\mu, m_{out}, \{0,1\}, (S^f_{m})_{m\in \overline{M}},
(\mathcal{R}^f_{m})_{m\in M})
$$
where each fuzzy multiset $S^f_{m}:V\times \{1\}\to \NN_{\infty}$
is defined from the corresponding crisp multiset $S_{m} : V\to
\NN_{\infty}$ in the natural way: for every $v\in V$,
$S^f_{m}(v,1)=S_{m}(v)$.  As far as the rules goes, each
$\mathcal{R}^f_{m}$ consists of the rules in $\mathcal{R}_{m}$ with
threshold mappings $\tau_{in}$ and $\tau_{out}$ that send, respectively,
every incoming and every outgoing reactive of the rule to 1 and all
other reactives to 0.

\begin{proposition}\label{lema}
Let $\Pi$ be a crisp P-system and $\Pi^{f}$ the fuzzy P-system defined
by it.  Then $Gen_{\Pi^{f}}=\chi_{Gen_{\Pi}}$, i.e., $Gen_{\Pi^{f}}(n)=1$
if $n\in Gen_{\Pi}$ and $Gen_{\Pi^{f}}(n)=0$ otherwise.
\end{proposition}

\proof{Since the configurations for $\Pi^f$ are $\{0,1\}$-valued, all
objects in the membranes and the environment in the initial
configuration of $\Pi^{f}$ are exact copies of the reactives in $V$
and only exact copies of reactives can enter the skin membrane from
the environment.  Therefore, at any moment of any computation with
respect to $\Pi^{f}$ every region defined by a membrane only contains
exact copies of reactives.  Furthermore, a rule $R\in \mathcal{R}_{m}$
can be applied to a configuration $(F_{m})_{m\in \overline{M}}$ of
$\Pi$ if and only if the corresponding rule $R^{f}$ in
$\mathcal{R}^f_{m}$ can be triggered in the corresponding
configuration $(F_{m}^{f})_{m\in \overline{M}}$ of $\Pi^{f}$ (the one
defined by $F_{m}^{f}(v,1)=F_{m}(v)$ for every $m\in \overline{M}$ and
$v \in V$), and the result of the (unique) application in $\Pi^{f}$ of
$R^{f}$ to $(F_{m}^{f})_{m\in
\overline{M}}$ is also the configuration corresponding to the result
of the application of $R$ in $\Pi$ to $(F_{m})_{m\in \overline{M}}$.

By the formal definitions of transition in crisp and fuzzy
P-systems, this argument also entails that every transition for $\Pi$ defines,
in a bijective way, a transition for $\Pi^f$, which produces the
configuration for $\Pi^f$ corresponding to the configuration produced
by the transition for $\Pi$.  Thus, every computation $C$ with respect
to $\Pi$ defines, also in a bijective way, a computation with respect
to $\Pi^{f}$, which we shall denote by $C^{f}$, in such a way that $C$
is halting if and only if $C^{f}$ is halting and, if they both are
halting, $H_{C^{f}}(1)=Out_{\Pi,C}$.

Therefore, for every halting computation $C$ with respect to $\Pi$,
$$
Out_{\Pi^{f},C^{f}}(n)=\left\{
\begin{array}{ll}
	1 & \mbox{ if $n=Out_{\Pi,C}$}\\
	0 & \mbox{ otherwise}
\end{array}
\right.
$$
and thus
$$
\begin{array}{rl}
Gen_{\Pi^{f}}(n) & =\left\{
\begin{array}{ll}
	1 & \mbox{ if there exists some halting computation $C^{f}$ w.r.t. 
	$\Pi^{f}$}\\
& \mbox{ such 
	that $Out_{\Pi^{f},C^{f}}(n)=1$}\\
	0 & \mbox{ otherwise}
\end{array}
\right.\\[3ex]
& =\left\{
\begin{array}{ll}
	1 & \mbox{ if there exists some halting computation $C$ w.r.t. 
	$\Pi$}\\
& \mbox{ such 
	that $Out_{\Pi,C}=n$}\\
	0 & \mbox{ otherwise}
\end{array}
\right.\\[3ex]
& =\left\{
\begin{array}{ll}
	1 & \mbox{ if $n\in Gen_{\Pi}$}\\
	0 & \mbox{ otherwise}
\end{array}
\right.
\end{array}
$$
as we claimed.}

This proposition remains true if
we enlarge the set $\{0,1\}$ in the definition of $\Pi^f$ to any
finite subset $I$ of $[0,1]$ containing $0$ and $1$ and then we set,
for every $m\in M$, $S^f_{m}(v,t)=S_{m}(v)$ if $t=1$ and
$S^f_{m}(v,t)=0$ otherwise, and $S^f_{env}(v,t)=S_{env}(v)$ for every
$t\in I^+$, but we still endow all rules with threshold functions that
take value 1 on every incoming or outgoing reactive.  In this case, at
any moment of any computation with respect to $\Pi^{f}$ every region
defined by a membrane (other than the environment) will only contain
exact copies of reactives and hence the proof of the last proposition
is still valid.

\section{Universality}

A \emph{fuzzy language} over an alphabet $\Sigma$ is a fuzzy subset
$L:\Sigma^{*}\to [0,1]$ of $\Sigma^{*}$.  Such a fuzzy language is
\emph{recursively enumerable}  when all its
levels
$$
L_{t}=\{w\in \Sigma^{*}\mid L(w)\geq t\},\quad t\in [0,1],
$$
are recursively enumerable in the usual sense; cf.\ \cite{fre}. Notice that, since
$L_{0}=\Sigma^{*}$, it is enough to consider in this definition the
$t$-levels with $t>0$.  Moreover, arguing as at the beginning of \S
2.1, we can see that if $L$ is $I$-valued, then it is enough to
consider its $t$-levels with $t\in I^+$.

Now, in parallel to the definition of a recursively enumerable subset of $\NN$ as the
set of lengths of some recursively enumerable language, we shall
say that a fuzzy subset $F:\NN\to [0,1]$ of $\NN$ is \emph{recursively
enumerable} when there exists some recursively enumerable fuzzy language
$L:\Sigma^{*}\to [0,1]$, over some alphabet $\Sigma$, such that, for every
$n\in \NN$,
$$
F(n)=\bigvee\{L(w)\mid w\in \Sigma^{*},\ |w|=n\}.
$$
Now we have the following characterization of  recursively enumerable fuzzy subsets of 
$\NN$ in terms of levels, which is the one we shall use henceforth.

\begin{proposition}
An $I$-valued fuzzy subset $F:\NN\to I$ is recursively enumerable if and only if  $F_{t}$ is 
a recursively enumerable subset of $\NN$, for every $t\in I^+$.
\end{proposition}

\proof{Let $L:\Sigma^{*}\to I$ be a  recursively enumerable fuzzy language such that, for 
every $n\in \NN$,
$$
F(n)=\bigvee\{L(w)\mid w\in \Sigma^{*},\ |w|=n\}.
$$
Then, for every $t\in I^+$,
$$
\begin{array}{rl}
F_{t}& =\{n\in\NN\mid \bigvee\{L(w)\mid w\in \Sigma^{*},\ |w|=n\}\geq 
t\}\\
&
=\{n\mid \mbox{there exists some $w\in \Sigma^*$ with $|w|=n$ such that 
$L(w)\geq t$}\}\\
& = \{|w|\mid w\in L_{t}\}
\end{array}
$$
is the set of lengths of a recursively enumerable language, and hence recursively enumerable itself. 

Conversely, let $F:\NN\to I$ be a fuzzy subset of $\NN$ such that each
$F_{t}$ is recursively enumerable, and consider the fuzzy language over a singleton
$\{a\}$
$$
\begin{array}{rrcl}
	L: & \{a\}^* & \to & I \\
	 & a^n & \mapsto & F(n)
\end{array}
$$
It is clear that $L_{t}=\{a^n\mid n\in F_{t}\}$ 
and hence, if every $F_{t}$ with $t\in I^+$ is recursively enumerable, the same happens
for each $L_{t}$ with $t\in I^+$.  Therefore, $L$ is a recursively enumerable fuzzy
language.  And it is also clear that, for every $n$,
$$
F(n)=L(a^n)=\bigvee\{L(w)\mid w\in \{a\}^*,\ |w|=n\},
$$
which finally implies that $F$ is recursively enumerable, too.}

Our goal now is to prove that a finite-valued fuzzy subset of $\NN$ is
recursively enumerable if and only if it is generated by a fuzzy P-system.  We begin
with the easy implication in this equivalence.

\begin{theorem}\label{facil}
	Every fuzzy subset of $\NN$ generated by a fuzzy P-system is recursively enumerable.
\end{theorem}

\proof{Let
$$
\Pi=(V, V_{out}, \mu, m_{out}, I, (S_{m})_{m\in \overline{M}},
(\mathcal{R}_{m})_{m\in M})
$$
be a fuzzy  P-system.  For every $t\in I^{+}$,
let $\Pi^{(t)}$ be the crisp P-system
$$
\Pi^{(t)}=(V\times I^{+}, V_{out}\times \{t\}, \mu, m_{out},
(S_{m})_{m\in \overline{M}}, (\mathcal{R}^{c}_{m})_{m\in M}),
$$
where each $S_{m}:V\times I^{+}\to \NN$ is now understood as a
multiset over $V\times I^{+}$ and, for every $m\in M$, the set of
rules $\mathcal{R}^{c}_{m}$ contains, for each
$((\underline{a},in;\underline{b},out),\tau_{in},\tau_{out})\in \mathcal{R}_{m}$, say
with $\underline{a}=a_{1}\ldots a_{p}$ and $\underline{b}=b_{1}\ldots
b_{q}$, each possible rule of the form
$$
\Bigl((a_{1},t_{i_{1}})\ldots (a_{p},t_{i_{p}}),in;
(b_{1},t_{j_{1}})\ldots (b_{q},t_{j_{q}}),out\Bigr)
$$
with $t_{i_{1}},\ldots,t_{i_{p}},t_{j_{1}},\ldots,t_{j_{q}}\in I^+$
such that $t_{i_{k}}\geq \tau_{in}(a_{k})$ for every $k=1,\ldots,p$
and $t_{j_{l}}\geq \tau_{out}(b_{l})$ for every $l=1,\ldots,q$, and it
only contains rules obtained in this way.  Notice thus that these
membrane systems $\Pi^{(t)}$ only differ in their sets of output
reactives, and hence they have exactly the same halting computations, 
but any such halting computation may produce in each $\Pi^{(t)}$ a 
different output.

Now, we can identify the configurations for $\Pi$ with the
configurations for each $\Pi^{(t)}$, by simply understanding a fuzzy
multiset over $V$ as a multiset over $V\times I^+$.  We can also
identify each application of a rule $R\in\mathcal{R}_{m}$ to a
configuration for $\Pi$ with the application of some rule contributed
by $R$ in $\mathcal{R}^c_{m}$ (a different rule for each application)
to the corresponding configuration for each $\Pi^{(t)}$.  From the
explicit description of transitions for crisp and fuzzy P-systems, we
deduce that we can actually identify each transition for $\Pi$ from a
configuration $(F_{m})_{m\in \overline{M}}$ to a configuration
$(\widehat{F}_{m})_{m\in \overline{M}}$ with a transition for each
$\Pi^{(t)}$ from the configuration corresponding to $(F_{m})_{m\in
\overline{M}}$ to the configuration corresponding to
$(\widehat{F}_{m})_{m\in \overline{M}}$.  This finally entails that
every halting computation with respect to any $\Pi^{(t)}$ corresponds
to a halting computation with respect to $\Pi$.

Now, let us fix an arbitrary $t_{0}\in I^{+}$; we want to prove that
the $t_{0}$-level $(Gen_{\Pi})_{t_{0}}$ is recursively enumerable.  For every halting
computation $C$ with respect to $\Pi$, let $H_{C}:I^{+}\to \NN$ be the
multiset on $I^{+}$ associated to it, and let $h_{C}^{(t_{0})}\in \NN$
be the output of the corresponding halting computation with respect to
$\Pi^{(t_{0})}$.  Since the output set of reactives of $\Pi^{(t_{0})}$
is $V_{out}\times \{t_{0}\}$, it is clear that
$h_{C}^{(t_{0})}=H_{C}(t_{0})$.

Now, by Lemma \ref{lemaprevi1}, we have that
$Gen_{\Pi}(n)\geq t_{0}$ if and only if $h_{C}^{(t)}= n$ for some 
halting computation $C$ and some $t\geq t_{0}$. Thus,
$n\in (Gen_{\Pi})_{t_{0}}$ if and only if $n\in Gen_{\Pi^{(t)}}$ 
for some
$t\geq t_{0}$, i.e.
$$
(Gen_{\Pi})_{t_{0}}=\bigcup_{t\geq t_{0}} Gen_{\Pi^{(t)}}.
$$ 
Since every $Gen_{\Pi^{(t)}}$ is a recursively enumerable subset of
$\NN$ and there are only a finite number of them, this implies that
$(Gen_{\Pi})_{t_{0}}$ is recursively enumerable, too.  And since $t_{0}$
was arbitrary, this shows that $Gen_{\Pi}$ is a
recursively enumerable fuzzy subset of $\NN$.}

The converse implication is given by the following result.

\begin{theorem}\label{univ}
Every recursively enumerable finite-valued fuzzy subset of $\NN$ is
generated by a fuzzy P-system.
\end{theorem}

\proof{Let $F:\NN\to I$ be a recursively enumerable $I$-valued fuzzy subset of $\NN$, with $I$
finite and containing 0 and 1.  Each level $F_{t}$, with $t\in I^+$,
is recursively enumerable, and therefore, by Theorem \ref{crisp}, it is generated by a
P-system
$$
\Pi^{(t)}=(V^{(t)},V_{out}^{(t)}, \mu, m_{out},  (S^{(t)}_{m})_{m\in
\overline{M}}, (\mathcal{R}^{(t)}_{m})_{m\in M})
$$
that satisfies the conditions listed in that theorem:
$V^{(t)}_{out}=V^{(t)}$; it only has symport rules; it has only two
membranes, the skin membrane $1$ and the output membrane $m_{out}=2$;
the only rules associated to the output membrane are
$(\alpha^{(t)},in)$, $(\#^{(t)},in)$ and $(\#^{(t)},out)$ for some
specific reactives $\alpha^{(t)}$ and $\#^{(t)}$; $\alpha^{(t)}$ is
the only reactive that may enter the output membrane in any halting
computation; and
$S^{(t)}_{1}(\alpha^{(t)})=S^{(t)}_{2}(\alpha^{(t)})=0$.  We shall
assume that the sets of reactives $V^{(t)}$ are pairwise disjoint.

Now, consider the fuzzy P-system
$$
\Pi=(V, V_{out}, \mu', m_{out}, I, (S_{m})_{m\in \overline{M}}, (\mathcal{R}_{m})_{m\in M})
$$
defined as follows:
\begin{itemize}
\item $V=\bigsqcup_{t\in I^+} V^{(t)}$.

\item $V_{out}=V$.

\item $\mu'$ is a linear tree obtained from $\mu$ by adding a third
membrane, labelled 3, as the new elementary membrane.

\item  $m_{out}=2$.

\item For every $m=1,2$, for every $v\in V$,
and for every $t,t'\in I^+$,
$$
S_{m}(v,t')=\left\{
\begin{array}{ll}
S_{m}^{(t)}(v) & \mbox{ if $v\in V^{(t)}$ and $t'=1$}\\
0 & \mbox{ otherwise}
\end{array}\right.
$$
And set $S_{3}(v,t)=0$ for every $v\in V$ and $t\in I^+$.

Thus, membranes 1 and 2 in $\Pi$ contain the sum of their contents in
all $\Pi^{(t)}$, with all objects being exact copies of the
corresponding reactives, while the new membrane 3 is empty at the
beginning.  Notice in particular that, for every $t,t'\in I^+$ and for
every $i=1,2,3$, $S_{i}(\alpha^{(t)},t')=0$.

Finally, as far as $S_{env}$ goes, if $v\in V^{(t)}$, then 
$S_{env}(v,t')=S_{env}^{(t)}(v)$ for every $t'\in
I^+$. I.e.,  the environment contains an unbounded number of copies of $v$
in $\Pi^{(t)}$ exactly when the environment in $\Pi$ contains an unbounded
number of copies of this reactive with any non-zero degree of
exactitude.

\item For every $m=1,2$, and for every rule $(\underline{a},in)$ or
$(\underline{a},out)$ in some $\mathcal{R}_{m}^{(t)}$, the set
$\mathcal{R}_{m}$ contains a corresponding rule
$$
((\underline{a},in),\tau) \mbox{ or } 
((\underline{a},out),\tau)
$$
where the threshold mapping $\tau$ is defined as follows: for every $v\in 
V$, if $|\underline{a}|_{v}=0$, then $\tau(v)=0$, and if 
$|\underline{a}|_{v}>0$, then 
$$
\tau(v)=
\left\{
\begin{array}{ll}
	t & \mbox{if $v=\alpha^{(t)}$ for 
some $t\in I^+$}\\
1 & \mbox{otherwise}
\end{array}
\right.
$$

On the other hand, if for every $t\in I^+-\{1\}$ we denote by $s(t)$
the least element in $I$ greater than $t$ (which exists because $I$ is
finite), then $\mathcal{R}_{3}$ contains, for every $t\in I^+-\{1\}$,
a rule
$$
((\alpha^{(t)},in),\tau)
$$
with $\tau(\alpha^{(t)})=s(t)$.

And $\mathcal{R}_{m}$, $m=1,2,3$, do not contain any rule other than
these ones.
\end{itemize}

To simplify the notations, set
$$
V_{a}=\{\alpha^{(t)}\mid t\in I^{+}\}.
$$

Thus, at the beginning, all membranes in $\Pi$ other than the environment only
contain exact copies of non-output reactives.  Furthermore, $\Pi$
contains rules of two types.  There are rules induced from
rules in some $\Pi^{(t)}$ that move exact copies of reactives in
$V-V_{a}$ as well as copies of reactives $\alpha^{(t)}$  with
degree of similarity at least $t$, in the same way as the
corresponding rule moved them in $\Pi^{(t)}$.  And 
rules that remove from the output membrane all copies of reactives
$\alpha^{(t)}$  with degree of similarity greater than the
corresponding $t$ and bury them in the elementary membrane.

In particular, no non-exact copy of a reactive in $V-V_{a}$ may ever
enter the skin membrane from the environment, and the only objects
similar to some $\alpha^{(t)}$ that enter it must have degree of accuracy
at least $t$.  Therefore, all objects that, at some moment of a
computation with respect to $\Pi$, are contained in some membrane
other than the environment, are either exact copies of reactives in
$V-V_{a}$ or similar to some $\alpha^{(t)}$ with degree of exactitude at
least $t$.  Moreover, no copy of a reactive $\alpha^{(t)}$ with degree of
similarity greater than $t$ may remain in the output membrane when a
computation halts.

Now, the fact that each rule involves only reactives in
some $V^{(t)}$ and the form of the rules in each
$\mathcal{R}^{(t)}_{2}$ imply that in a given configuration for $\Pi$,
the application conditions for a rule coming from $\Pi^{(t)}$, a rule
coming from $\Pi^{(t')}$ with $t\neq t'$ and any one of the new rules
in $\RR_{3}$ are independent of each other.

Using these remarks, one can easily see that every transition with
respect to $\Pi$ consists of the application in parallel of families
of rules coming from rules that form transitions with respect to some
P-systems $\Pi^{(t)}$ plus the application of all rules in $\RR_{3}$
necessary to remove from the output membrane all copies of reactives
$\alpha^{(t)}$ with degree of exactitude greater than $t$.  Therefore,
every halting computation with respect to $\Pi$ corresponds to a
family of halting computations $(C_{t})_{t\in I^+}$ performed in
parallel, every $C_{t}$ with respect to the corresponding $\Pi^{(t)}$.
These computations $C_{t}$ are uniquely determined by $C$, and they
may halt at different moments: $C$ halts when all computations $C_{t}$
halt and no copy of any $\alpha^{(t)}$ with degree of exactitude greater
than $t$ remains in the output membrane.

Then, the output component $H(C)_{2}$ of the halting configuration of
a halting computation $C$ corresponding to a family of halting
computations $(C_{t})_{t\in I^+}$ with respect to the P-systems
$(\Pi^{(t)})_{t\in I^+}$ satisfies that, for every $t,t'\in I^+$,
$$
H(C)_{2}(v,t)=\left\{
\begin{array}{ll}
(H_{C_{t}})_{2}(\alpha^{(t)})& \mbox{ if $v=\alpha^{(t)}$}\\
0 & \mbox{ otherwise}
\end{array}
\right.
$$
Hence, the mapping $H_{C}$ associated to this halting computation $C$ 
is given by 
$$
H_{C}(t)=(H_{C_{t}})_{2}(\alpha^{(t)})=Out_{\Pi^{(t)},C_{t}}.
$$

Now, by Lemma \ref{lemaprevi1}, for every $t_{0}\in I^+$,
$Gen_{\Pi}(n)\geq t_{0}$ if and only if there exists some halting
computation $C$ with respect to $\Pi$, corresponding to some family of
halting computations $(C_{t})_{t\in I^+}$, and some $t'\geq t_{0}$ such
that $H_{C}(t')=n$.  Since, moreover, every halting computation with
respect to any $\Pi^{(t')}$ will be part of some halting computation
with respect to $\Pi$, this condition is equivalent to the existence
of some $t'\geq t_{0}$ and some halting computation $C_{t'}$ with
respect to $\Pi^{(t')}$ such that $Out_{\Pi^{(t')},C_{t'}}=n$, i.e., 
to the fact that $n\in Gen_{\Pi^{(t')}}$ for some $t'\geq t_{0}$.  In all,
this shows that 
$$
(Gen_{\Pi})_{t_{0}}=\bigcup_{t'\geq t_{0}} 
Gen(\Pi^{(t')})=\bigcup_{t'\geq t_{0}} F_{t'}.
$$ 
And finally, since the $t$-levels of $F$ are decreasing in $t$,
this entails that 
$$
(Gen_{\Pi})_{t_{0}}=F_{t_{0}}.
$$
Thus, $Gen_{\Pi}$ and $F$ have exactly the same levels, 
and therefore they are the same fuzzy subset of $\NN$.}

\section{Conclusion}

In this paper we have introduced a fuzzy version of the
symport/antiport model of membrane systems that uses inexact copies of
reactives in the transitions.  Then, we have proved that this fuzzy
model of computation is universal, in the sense that it generates all
recursively enumerable finite-valued fuzzy subsets of $\NN$.  This
means a first step towards the use of fuzzy methods to answer a
question posed by Gh.\ P\u{a}un in the last problem of his first list
of open problems in membrane computing \cite{probl1}: ``What about
`approximate' computing, whatever this can mean?''

The key ingredients in our model are the use of fuzzy multisets in
configurations, the endowment of evolution rules with threshold
mappings that determine the degree of exactitude of objects to
reactives in order to be affected by the rules, and an appropriate way
of evaluating the content of the output membrane at the end of a
halting computation.  These ingredients could also be used
\textsl{mutatis mutandis} to define the fuzzy version of any other
membrane computing model: in this paper we only considered the
symport/antiport model for simplicity.

When proving the universality of our fuzzy P-systems we have not
addressed any minimality question like the least number of membranes
or the least number of output reactives that are necessary to generate
all recursively enumerable finite-valued fuzzy subsets of $\NN$: we leave these
 as open problems.  Nevertheless, let us mention here that we
have specified a set of output reactives in our P-systems with the
only purpose of simplifying the proof of Theorem \ref{facil}.  Nowhere
else in this paper it is needed and, as one would expect, that theorem
can also be proved without distinguishing output reactives, but at the
prize of using more involved crisp P-systems: this proof will appear
elsewhere \cite{Moya}.

We would like to point out here that the finite-valuedness of the
fuzzy subsets of $\NN$ generated by our fuzzy P-systems is not due to
the specification beforehand of the finite set of possible values $I$,
but rather to the finiteness of the sets of rules and the initial
configuration.  This effect also appears for instance in fuzzy
grammars where only a finite number of rules have a non-zero weight.

Indeed, assume for the moment that such an $I$ is not
specified and that in all fuzzy subsets and multisets used in the
definition of a fuzzy P-system $\Pi$, as well as in the description of
how it works, all sets $I$ and $I^+$ are replaced by $[0,1]$ and
$]0,1]$, respectively.

Since the ensemble of evolution rules $\bigcup_{m\in
M}\mathcal{R}_{m}$ in $\Pi$ is finite and each rule only involves a
finite set of reactives, it is clear that the set of the images of all
threshold functions of all rules is a finite subset of $[0,1]$.  
Moreover, similarly to the crisp case, we would assume that
the initial content of regions other than the environment is finite
(in our definition it is entailed by the fact that $I$ is finite), and
thus we would impose that the initial configuration is given by
finite-valued fuzzy multisets $S_{m}$ for every $m\in M$ (but not for
the environment).  Hence, the set of possible exactitude values of
objects to reactives in the initial contents of the regions $K_{m}$
described by these multisets form a finite subset of $[0,1]$.  Let the
union of these two finite sets be $J=\{t_{0},t_{1},\ldots,t_{m}\}$
with $t_{0}<t_{1}<\ldots <t_{m}$, and assume for simplicity that
$t_{0}=0$, $t_{m}=1$.  To the effect of triggering a rule, any two
copies of the same reactive with degrees of accuracy $t,t'$ in some
interval $[t_{l},t_{l+1}[$, $l=0,\ldots,m-1$, are indistinguishable, and
any object with degree of accuracy $t\in ]t_{l},t_{l+1}[$ to some
reactive must come from the environment.

Additionally, we still must impose that the environment contains
an unbounded homogeneous supply of some reactives. In order not to
distinguish any degree of similarity, we would impose it by assuming
that $S_{env}(v,t)$ is either 0 for every $t\in ]0,1]$ or $\infty$ for
every $t\in ]0,1]$.

This would entail that, for every $t,t'\in ]t_{l},t_{l+1}[$,
$l=0,\ldots,m-1$, and for every computation $C$ with respect to $\Pi$,
there exists a computation $C'$ such that $H_{C'}(t')=H_{C}(t)$. 
Indeed, $C'$
has the same ordered sequence of families of rules as $C$, but for
every $v$ and every application of a rule $R$, $enter^{R}_{v}(t')$ and
$exit^{R}_{v}(t')$ in the application in $C'$ take the values of
$enter^{R}_{v}(t)$ and $exit^{R}_{v}(t)$ in the corresponding
application in $C$, and $enter^{R}_{v}(t)$ and $exit^{R}_{v}(t)$ in the
application in $C'$ take the values of $enter^{R}_{v}(t')$ and
$exit^{R}_{v}(t')$ in the corresponding application in $C$.

Therefore, if some $t\in ]t_{l},t_{l+1}[$ is contained in 
$$
\{t\in ]0,1]\mid H_{C}(t)= n \mbox{ for some halting computation $C$}\} 
$$
for some $n\in \NN$, then the whole interval $[t_{l},t_{l+1}[$ is
contained in this set.  This entails that the supremum of this set,
which would define $Gen_{\Pi}(n)$, belongs to $J$.  Thus, after all,
the fuzzy set $Gen_{\Pi}$ is still finite-valued, and moreover its set 
of values is contained in  $J$.

Besides, $S_{env}(v,t)=\infty$ for some $v\in V$ and every $t\in
]0,1]$ entails that the universe in non-countably infinite, and that a
non-countably infinite number of computations may exist.

These observations, and the obvious fact that working with fuzzy sets
and multisets that are explicitly specified as finite-valued greatly
simplifies  all notations, definitions and proofs, motivated us
to restrict ourselves from the very beginning to the finite set $J$,
or rather a finite extension $I$ of it, as the set of values of any
fuzzy set related to the fuzzy P-system $\Pi$.

To end this paper, we would like to point out that, although formally
correct, our specific approach has a drawback from the fuzzy
mathematics point of view. The association to a multiset $H:I^+\to 
\NN$ of the fuzzy subset of $\NN$
$$
\begin{array}{rrcl}
\CC(H): & \NN & \to & I\\
	& n & \mapsto & \bigvee\{t\mid H(t)= n\}
\end{array}
$$
that underlies our definition of the output of a halting computation
with respect to a fuzzy P-system is not additive in any natural sense,
and in particular it  cannot be considered a fuzzy
cardinality; see \cite{CR-fb}. We have tried to use some specific simple
fuzzy cardinalities in this step, and we have obtained that the
resulting fuzzy P-systems did not generate all finite-valued recursively enumerable
fuzzy subsets of $\NN$, but we have not ruled out the possibility of
using some other, cunningly chosen, fuzzy cardinality.  Our current
research agenda includes this problem, as well as the problem of
getting rid of the assumption used in this paper that an object can
only be similar to one reactive.
\smallskip

\textbf{Acknowledgments.} This work has been partially supported by the
Spanish DGES and the EU program FEDER, project BFM2003-00771.

\end{document}